\documentclass[11pt]{article}

\usepackage[preprint]{acl}
\usepackage{booktabs}
\usepackage{times}
\usepackage{natbib}
\usepackage{latexsym}
\usepackage{stfloats}
\usepackage[most]{tcolorbox}
\usepackage[T1]{fontenc}

\usepackage[utf8]{inputenc}

\usepackage{microtype}

\usepackage{inconsolata}

\usepackage{graphicx}
\usepackage{amsmath}
\usepackage{float}
\usepackage{algorithm}
\usepackage{algpseudocode}
\usepackage{multirow}
\title{EmpiriGraph-Psy: A Dataset and LLM Pipeline for Extracting Empirical Relation Graphs from Psychology Abstracts}

\author{
Danqin Zhao$^{\alpha,*}$ \quad
Yicun Liu$^{\beta,*}$ \quad
Xingwei Tan$^{\gamma}$ \quad
Thomas T. Hills$^{\alpha}$ \\
$^{\alpha}$ Department of Psychology, University of Warwick \\
$^{\beta}$ Mathematical Sciences Institute, The Australian National University\\
$^{\gamma}$ School of Computer Science, University of Sheffield\\
\texttt{\{Danqin.Zhao,T.T.Hills\}@warwick.ac.uk} \\
\texttt{u7579143@anu.edu.au} \\
\texttt{Xingwei.Tan@sheffield.ac.uk} \\
$^{*}$ Equal contribution.
}

\vspace{20mm}

\begin{document}
\maketitle
\begin{abstract}
Existing scientific relation extraction benchmarks mainly target domains such as computer science, where entities are tasks, methods, datasets, materials, or metrics. This leaves a gap in variable-oriented empirical fields such as psychology, where findings are expressed as relations among constructs, measurements, interventions, and outcomes. We introduce variable-centered empirical graph extraction, the task of mapping scientific abstracts to typed graphs whose nodes are normalized variables and whose edges represent empirical and hierarchical relations. To support this task, we construct EmpiriGraph-Psy, a benchmark of 210 psychology abstracts annotated by domain-trained annotators with normalized variables, concept hierarchies, empirical relation types, and validation states. We evaluate frontier and open-weight LLMs using both direct extraction and a staged graph-construction pipeline that separates variable extraction, normalization, hierarchy construction, evidence selection, relation extraction, and edge validation. The staged pipeline substantially outperforms direct extraction, with the best configuration achieving a macro-F1 of 0.74. Error analysis shows that moderation relations and concept hierarchies remain the most challenging cases, highlighting the difficulty of extracting higher-order empirical claims and implicit abstraction structure from scientific abstracts.~\footnote{Experimental code: \url{https://github.com/foxxis-dq828/EmpiriGraph-Psy}}
\end{abstract}

\section{Introduction}
Scientific relation extraction aims to identify concepts and relations from unstructured research text and represent them as structured graphs. Existing benchmarks have largely focused on computer science and NLP papers (e.g., \citealt{Gabor_2018_SemEval}) , where scientific entities are typically tasks, datasets, models, and metrics. While these schemas support model comparison and evaluation, they are less suited to variable-oriented empirical fields such as psychology, social science, and health research. In these disciplines, knowledge is often organized around variables and their empirical relations, including covariation, intervention--effects, mechanisms, and contextual conditions.


In this paper, we construct a corpus of psychology abstracts because psychology is a representative variable-oriented empirical field: its findings are commonly expressed as relations among such variable relations. For example, abstracts may report associations between leadership and personality (e.g., \citealt{ANDERSEN20061078}), effects of psychological interventions on patient outcomes (e.g., \citealt{Anderson2018}), or moderation by family environment in the relation between genetic risk and behavior (e.g., \citealt{Cadoret1995}). Extracting such relations can support the construction of variable-centered knowledge graphs for a broad range of empirical disciplines. It further enables large-scale evidence synthesis and historical analyses of how variables and theories emerge, stabilize, change, or disappear across the literature.

\begin{figure*}[ht]
    \centering
    \includegraphics[width=0.7\textwidth, keepaspectratio]{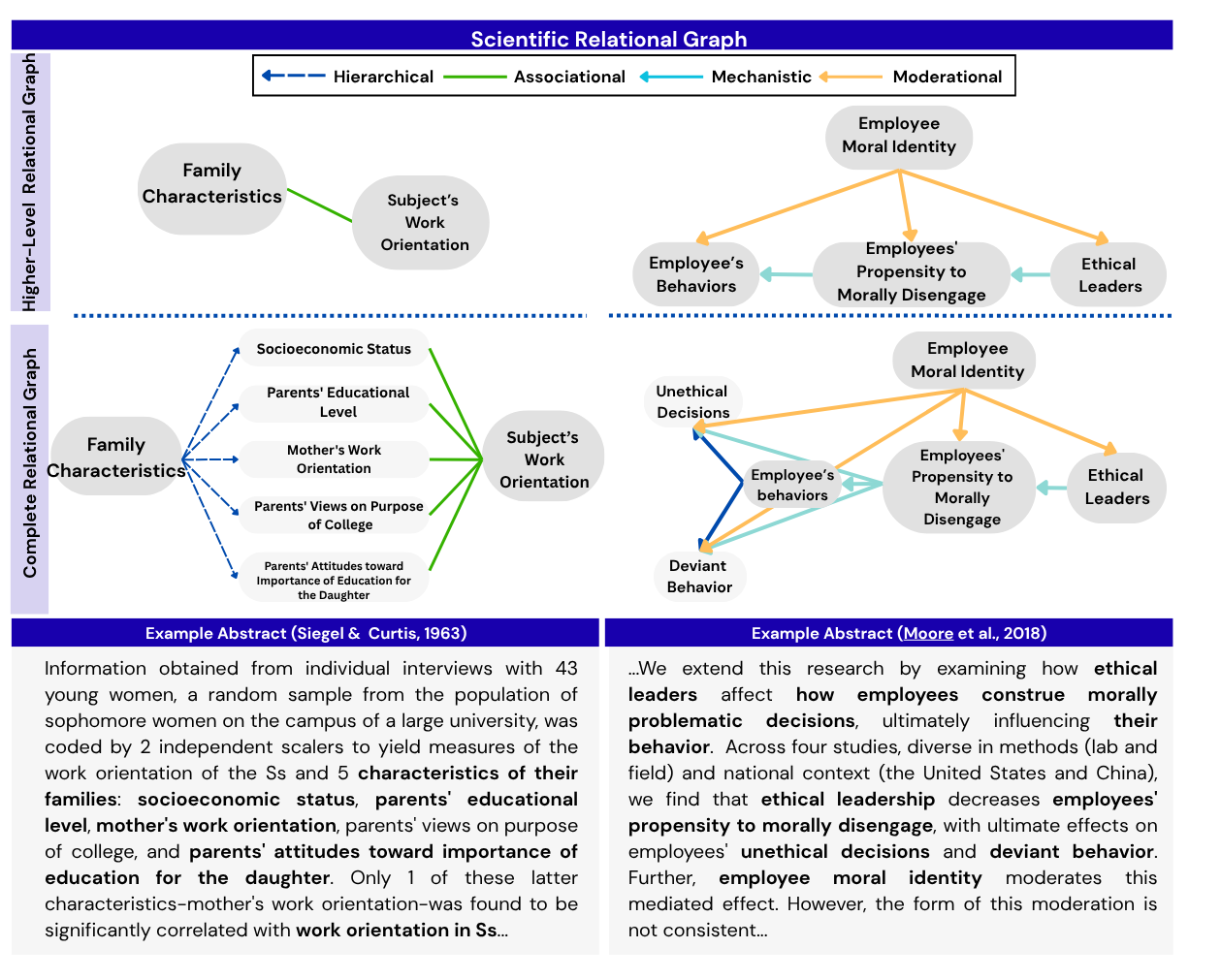}
    \caption{Illustration of the variable-centered relational graph extraction task. The input abstract is transformed into a typed variable graph, where the upper layer represents higher-level constructs and the lower layer represents finer-grained variables or dimensions. Edges capture associational, mechanistic, moderational, and hierarchical relations.}
    \label{fig:intro_figure}
\end{figure*}

Variable-centered graph extraction poses challenges beyond standard entity and relation extraction. First, variable mentions require semantic normalization: the same construct may be expressed through synonyms, abbreviations, measurement instruments, or theoretically related terms. Second, empirical findings are often stated at multiple levels of abstraction. An abstract may relate broad constructs while also specifying relations among their finer-grained dimensions, as illustrated in Figure \ref{fig:intro_figure}. Flattening these levels loses theoretical structure, while treating them as unrelated variables fragments the evidence. Third, relation classification requires contextual reasoning, including distinguishing associational, directional, mechanistic, and moderational claims, and identifying whether a relation is validated, hypothesized, or null. These challenges call for an extraction framework that jointly models variable identification, normalization, hierarchy, evidence grounding, and relation classification.

We address these challenges with a multi-stage LLM pipeline that decomposes graph construction into variable extraction, semantic normalization, hierarchy construction, evidence selection, relation extraction, and edge validation. This decomposition follows the structure of the annotation task: variables are first identified and canonicalized, hierarchical edges explicitly encode abstraction structure, and empirical edges are classified as associational, mechanistic, or moderational with their directionality and validation status.


To evaluate predicted graphs, we use a structure-first alignment procedure inspired by maximum common subgraph matching. Rather than relying on surface overlap between variable names, this evaluation aligns predicted and gold graphs under partial node matching and measures whether associational, mechanistic, moderational, and hierarchical edges are recovered with the correct validation status. This allows us to separate structural graph recovery from surface variation in variable naming.


Our contributions are as follows:
\begin{itemize}
    \item We introduce Empirical Research Knowledge Graph Extraction, a variable-centered task for mapping empirical research abstracts into typed graphs over normalized variables.
    \item We construct EmpiriGraph-Psy, a benchmark of 210 psychology abstracts annotated with normalized variables, validation states, and associational, mechanistic, moderational, and hierarchical relations.
    \item We propose a multi-stage LLM pipeline for Empirical Research Knowledge Graph Extraction and show that it substantially outperforms direct prompting across multiple LLMs.
    \item We introduce a structure-first graph evaluation framework that aligns predicted and gold graphs under partial node matching and measures typed edge recovery.
\end{itemize}

\section{Related Work}

\subsection{Relation Extraction from Scientific Materials} 
Recent studies have extensively explored the role of NLP techniques in extracting, organizing, and synthesizing scientific findings from large-scale academic corpora\citep{wang2020microsoft, chen2021capturin,li2024scitopic,katz2024llm}. As noted by \citet{zhao_2024_review_on_relational_extraction}, the number of scientific publications has grown exponentially, making it increasingly infeasible for researchers to manually discover and synthesize important scientific facts embedded in unstructured text. In this context, relation extraction (RE), which aims to automatically identify structured relations among scientific entities, serves as an important infrastructure for the science of science, supporting applications such as scientific knowledge graph construction, text summarization, and scientific leaderboard construction\citep{Luan_2018_Multi_Task_Identification, Mondal_2021_NLP_Knowledge_Graph, Dagdelen_2024_structure_extract_chemical_LLM,Hou_2019_leaderboard, Sahinuc_2024_LLM_Leaderboard}.

Early scientific relation extraction studies can be traced to SEMEVAL-2017\citep{Augenstein_2017_extracting_keyphrases_and_relations,Luan_2017_scientific_relational_extraction}, which introduces a cross-domain scientific keyphrase tagging task, identifying TASK, PROCESS and MATERIAL spans across Computer Science, Materials Science, and Physics to capture domain-specific scientific concepts from text. Later benchmarks expanded scientific relation extraction with richer schemas for scientific entities and relations, including fine-grained relation types in SemEval-2018 Task 7 and joint entity--relation--coreference annotation in SciERC \citep{Gabor_2018_SemEval,Luan_2018_Multi_Task_Identification}. More recently, SciNLP extends scientific information extraction from abstracts to full-text NLP papers, annotating 60 ACL long papers with fine-grained \textit{task}, \textit{method}, \textit{dataset}, and \textit{metric} entities and 11 relation types, including \textit{UsedFor}, \textit{EvaluatedOn}, \textit{MeasuredBy}, and \textit{CompareWith} \citep{Duan_sciNLP_2025}.

Despite the progression of scientific information extraction, existing benchmarks show that entity recognition and type classification are comparatively tractable, while relation discovery remains the main bottleneck. In SemEval-2018 Task 7, systems perform better when candidate relation instances are given, but drop when they must both identify and classify relations from raw abstracts \citep{Gabor_2018_SemEval}. This gap is even more pronounced in full-text settings, where related entities may be separated by longer contexts \citep{Duan_sciNLP_2025}. Meanwhile, existing studies on scientific relation extraction mostly focus on NLP or computer science papers, with a smaller line of work considering other domains. For example, biomedical relation extraction (Bio-RE) identifies relations among biomedical entities, such as chemical--disease and gene--disease relations \citep{Shang_2025_biomedical_extraction}, while chemistry and materials science studies extract chemical reaction data \citep{Guo_2022_automated_chemical_extraction} and material synthesis procedures \citep{Yang_2022_PcMSP}.

\subsection{LLM in Relation Extraction} 
LLMs exhibit in-context learning \citep{Min_2022} and achieve promising results across many NLP tasks such as text classification, fact retrieval, and natural language inference  \citep{brown2020language, wei2022finetuned}. Recent studies further show that LLMs exhibit competitive performance on relation extraction tasks under instruction-tuned settings, in some cases approaching or surpassing standard fully supervised methods \cite{Wan_2023_GPT_RE, wadhwa2023revisiting, Tan_2025_salient_event}. LLM has also widely applied in scientific relation extraction across different disciplines including materials science \citep{Dagdelen_2024_structure_extract_chemical_LLM,Foppiano2024MiningExperimentalData} and biomedicine, where they facilitate structured knowledge acquisition from scientific literature \citep{Shang_2025_biomedical_extraction,Laskar_2025_biomedical_extraction_LLM}. Specifically, Chain-of-Thought (CoT) prompting has emerged as an effective mechanism for improving reasoning-intensive NLP tasks by eliciting intermediate reasoning steps from LLMs \citep{wei2022chain,wadhwa2023revisiting}. It improves relation extraction performance by enhancing relation grounding and aligning dispersed textual evidence with candidate relations, thereby mitigating the aforementioned limitations of conventional supervised approaches in relation discovery \citep{Ma_2023_CoT_ER, wadhwa2023revisiting}. These findings suggest that LLMs provide a promising approach for extracting empirical relations which requires contextual understanding and nuanced semantic reasoning. Despite these strengths, no existing LLMs relation extraction system focus on scientific abstraction, and adapting LLMs to this task is non-trivial. We thus apply LLMs to extract empirical research knowledge graphs from scientific abstracts.

\section{Background}
In this section, we define the task of Empirical Research Knowledge Graph Extraction.
Given a scientific text document $X$ containing $n$ tokens, the goal of the task is to create a graph 
$G = (V, E)$, where $V$ denotes a set of vertices representing scientific variables or constructs and $E$ denotes a set of edges representing the relationships between the variables. We distinguish between empirical relation edges and conceptual edges. Empirical edges describe substantive relationships among variables, including associational, mechanistic, and conditional relations. Conceptual edges depict abstraction relations between higher-level constructs and lower-level variables, dimensions, or indicators.
An edge is denoted as $(v_i, r, v_j)$, where $v_i$ is the head node, $v_j$ is the tail node, and $r$ is the relation between the two nodes.
Relation $r\in \mathcal{R}$ is assigned with one of the categories in the following label set:
\begin{equation}
\begin{aligned}
\mathcal{R} = \{&
\text{Associational}, \text{Mechanistic}, \\
&
\text{Moderational}, \text{Hierarchical}\}.
\end{aligned}
\end{equation}

An associational edge describes that the head node $v_i$ and the tail node $v_j$ covary or are correlated, without specifying a causal or mechanistic relationship between them, such as a correlation, covariation, or a group difference associated with $v_i$, without claiming that $v_i$ mechanistically affects $v_j$.

A mechanistic edge is annotated when the text further explains the mechanism underlying the covariance between $v_i$ and $v_j$, such that variable $v_i$ affects, predicts, influences, enables, or otherwise has a directional effect on variable $v_j$.

A conditional edge is annotated when a third variable $v_k$ conditions an established relationship between variables, such as $v_i \rightarrow v_j$, by altering its strength, direction, or statistical significance. This edge type primarily captures moderation or interaction effects.A conditional relation in which $v_k$ moderates the relationship $v_i \rightarrow v_j$ is encoded in the knowledge graph as $(v_k, Conditional, v_i)$ and $(v_k, Conditional, v_j)$.

For each empirical edge, we further annotate its validation state, using three labels: \textit{validated}, \textit{null}, and \textit{hypothesized}. An edge is labeled as \textit{validated} when the text reports that the relationship is supported by the study's empirical results. An edge is labeled as \textit{null} when the relationship is tested but not supported, such as when the abstract reports an insignificant effect. An edge is labeled as \textit{hypothesized} when the relationship is proposed as a hypothesis or expected, but the text does not report empirical evidence confirming or rejecting it.

A major annotation ambiguity arises because abstracts often reports variable relationships at different levels of abstraction. For example, as illustrated in Figure ~\ref{fig:intro_figure} (b), the abstract first describes how \textit{ethical leader} indirectly influence \textit{employee's behavior}, and then specifies two concrete forms of \textit{employee's behavior} -- \textit{unethical decision} and \textit{deviant behavior}. In this case, \textit{employees' behavior} is annotated as a higher-level construct, representing the main research objects, while \textit{unethical decision} and \textit{deviant behavior} are annotated as lower-level variables or specific behavioral outcomes. 

We introduce hierarchy edges for representing such abstraction relations. A hierarchy edge connects a higher-level construct to a lower-level variable. The former includes constructs, theoretical categories, or research objects, while the latter denotes a more specific dimension, indicator, or measurement of that construct. The lower-level graph retains all specific relationships between different dimensions of a construct or specific measurements. The higher-level graph collapses lower-level variables into their corresponding higher-level constructs, yielding a more abstract representation of the theoretical relationships among constructs. Such multi-level coding allows for different levels of analytic purposes. For downstream projects such as synthesizing research findings into a knowledge graph, higher-level relational graphs provide descriptions of main research objects and relationships. While for analyzing historical changes in scientific findings, a complete relational graph will retain the organization of relationships in the abstract.

\section{Dataset and Human Annotation}
\subsection{Abstract Corpus Construction}

We collect a corpus of abstracts from psychology journals. Since the frequency, density, and linguistic realization of different relation types may vary substantially across historical periods, we designed a stratified sample with broad temporal coverage rather than sampling only from recent publications. Our dataset covers abstracts from six psychology journals with long publication histories, relatively high 5-year impact factors, and represents different subfields of psychology. We include only original author-written abstracts and exclude retrospectively added machine-generated abstracts. 

\subsection{Data Collection and Validation}
\label{sec:data-collection}
Three annotators participated in the project. All annotators were psychology students at either the PhD or undergraduate level. Annotations were conducted using a customized annotation platform built on top of Label Studio. The details can be found in Appendix~\ref{sec:appendix1}. Prior to the formal annotation phase, all coders received training on the coding scheme and completed a qualification task consisting of 10 abstracts. These materials were used only for training and were excluded from the final coding dataset. After coders passed this initial training stage, they proceeded to the formal annotation task. During annotation, the coding guidelines were iteratively discussed and refined through team discussions.

Gold graphs were constructed through a two-stage process. First, the three annotators jointly covered the full corpus of 210 abstracts, with a 50-abstract subset independently annotated by all three coders for reliability assessment. Second, after agreement analysis, the annotations were reviewed by the annotator team. Disagreements in variable boundaries, variable normalization, hierarchy edges, relation types, and validation-state labels were discussed and resolved. The resulting reviewed annotations were used as the final gold graphs for model evaluation.

\begin{figure*}[ht]
    \centering
    \includegraphics[width=0.8\textwidth, keepaspectratio]{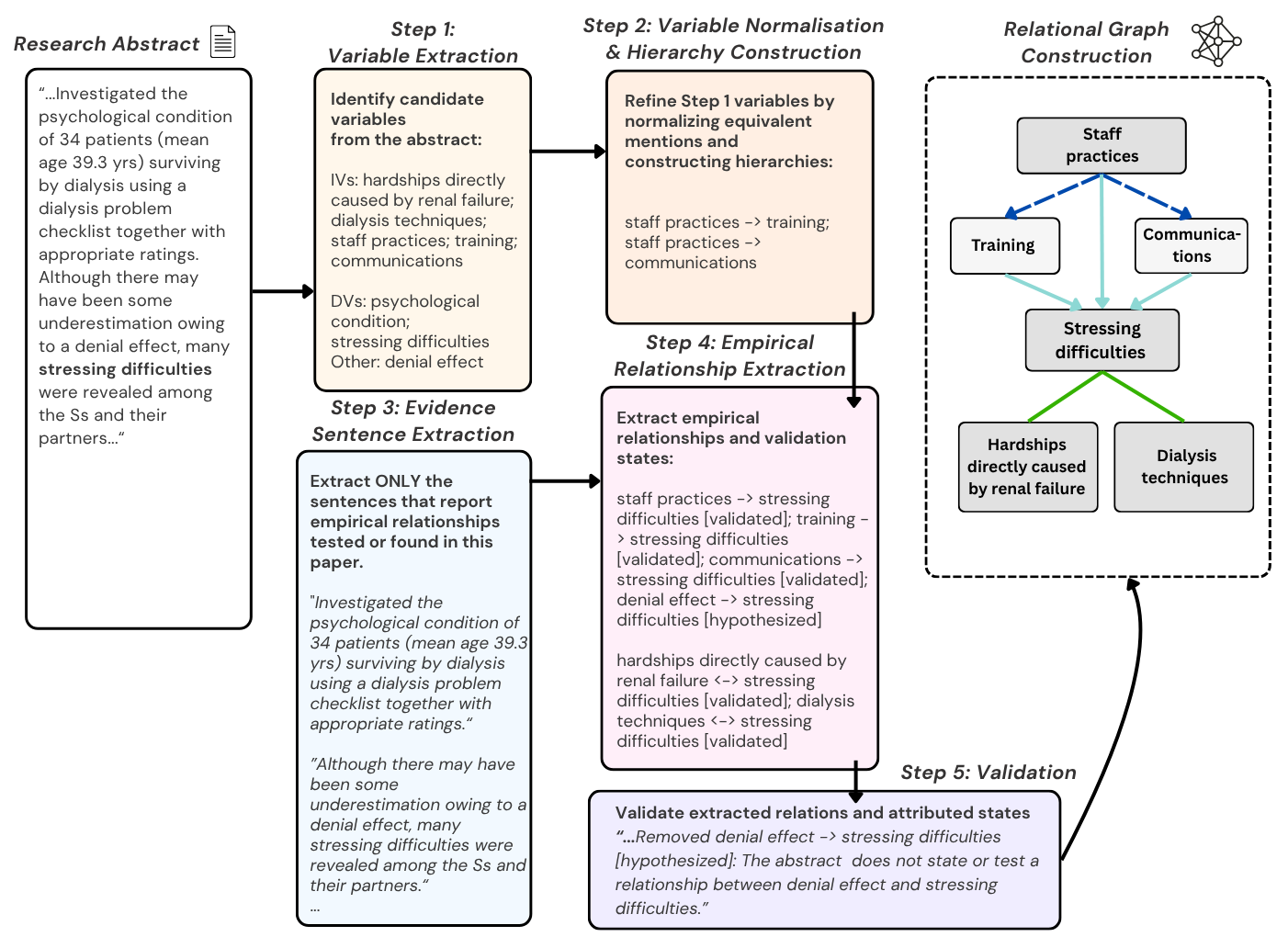}
    \caption{Overview of the proposed variable-centered relational graph extraction pipeline. The system maps a scientific abstract into a structured variable graph through variable extraction, normalization and hierarchy construction, evidence sentence extraction, graph construction, and edge validation.}
    \label{fig:method_figure}
\end{figure*} 

\section{Methodology}

\subsection{Pipeline}
Figure \ref{fig:method_figure} shows the overall structure of the proposed pipeline. We decompose graph construction into five sequential subtasks: variable extraction, variable normalization and hierarchy construction, evidence sentence extraction, empirical relationship extraction, and validation. Each stage produces a structured intermediate output, which is then passed to the next stage as context. Below we summarize the role of each stage.

\begin{tcolorbox}[
    breakable,
    colback=green!5,
    colframe=green!60!black,
    title=Deductive Task
]
\small

\textbf{Step 1: Variable Extraction}

In Step 1, the model identifies candidate variables from the abstract and proposes normalized variable spans/names. This stage emphasizes variable identification because downstream stages cannot recover variables that are never extracted.

\vspace{0.5em}

\textbf{Step 2: Variable Normalization and Hierarchy Construction}

In Step 2, the system builds a variable hierarchy and performs canonicalization. Practically, this stage merges near-duplicate mentions, links higher-level and lower-level variables, and enforces a consistent variable vocabulary for later relation extraction.

\vspace{0.5em}

\textbf{Step 3: Evidence Sentence Extraction}

In Step 3, the model selects evidence-bearing sentences likely to contain relational information with reference of the variable list generated from Step 2. This acts as an information bottleneck that improves precision by reducing distracting context and forcing the model to justify edge decisions from explicit textual evidence. 

\vspace{0.5em}

\textbf{Step 4: Graph Construction}

In Step 4, the model predicts and classifies empirical relations among variables (Associational, Mechanistic, Moderational), producing the initial graph structure. 

\vspace{0.5em}

\textbf{Step 5: Edge Validation}

In Step 5, the system re-validates extracted edges and relation attributes, correcting likely false positives and improving accuracy. This final pass functions as a consistency and quality-control stage before evaluation.

\end{tcolorbox}


\subsection{Graph Evaluation}
\label{sec:graph-eval}

Exact node-label matching is unsuitable because the same scientific variable may
be expressed by different surface forms across annotators or models. We therefore
evaluate graphs using a structure-first alignment. The gold graph G refers to the finalized human-annotated graph produced through the procedure described in Section \ref{sec:data-collection}. Let the gold graph be
\(G=(V_G,E_G)\) and the predicted graph be \(P=(V_P,E_P)\), where each edge is
directed and typed:
\begin{equation}
    e=(u,v,\tau),\tau\in\mathcal{T}.
\end{equation}

We search for an injective partial mapping
\begin{equation}
\phi:V_G\rightarrow V_P\cup\{\emptyset\},
\end{equation}
where \(\emptyset\) denotes an unmatched gold node. A gold edge
\((u,v,\tau)\in E_G\) is counted as matched if
\begin{equation}
    \phi(u)\neq\emptyset,\phi(v)\neq\emptyset,(\phi(u),\phi(v),\tau)\in E_P.
\end{equation}

The alignment is chosen to maximize typed edge overlap:

\begin{equation}
\footnotesize
    \phi^\star=\arg\max_{\phi}\left|\{(u,v,\tau)\in E_G:(\phi(u),\phi(v),\tau)\in E_P\}\right|.
\end{equation}

Let \(m_\star\) be the number of matched edges under \(\phi^\star\). We then compute precision, recall, and F1 score based on $m_\star$, $|E_P|$.

We report three complementary views: \(M_{\text{typed}}\), computed on the full
directed typed graphs; \(M_{\text{higher}}\), computed after projecting both
graphs onto higher-level nodes; and \(M_{\text{agnostic}}\), computed after
collapsing all edge types into a single undirected relation type. Optimization
details, preprocessing, per-type scores, and complexity analysis are provided in
Appendix~\ref{app:graph-eval}.

Because the goal of this paper is graph extraction rather than node-label normalization, we treat node alignment as an auxiliary component of graph-level evaluation. Nevertheless, since the structure-first metric may align nodes with different surface forms, we further validate the aligned node pairs in Appendix~\ref{app:node-validation}. Across all aligned node pairs, the mean embedding-based cosine similarity is $0.735$. Manual inspection of $100$ stratified randomly sampled pairs found that $87$ pairs referred to the same variable or construct. This suggests that most structure-aligned node pairs are semantically valid, although a minority of alignments remain noisy.

\section{Experiment}
\subsection{Baseline}
We first constructed a direct-prompting baseline by providing GPT-5.4 with a general task description, edge-type definitions, and the desired output format, and asking it to generate the complete graph in a single step. To evaluate whether explicit task decomposition improves Empirical Research Knowledge Graph Extraction, we compared three GPT-5.4-based settings: direct one-step prompting, a collapsed pipeline prompt that describes all five stages within a single request, and the full staged pipeline in which the five stages are executed as separate requests.

We then evaluated the full staged pipeline across several high-performing large language models selected with reference to the LLM leaderboard  reasoning index ranking\footnote{https://llm-stats.com/}: GPT-5.2, GPT-5.4, Claude Sonnet 4.6, Claude Opus 4.7, DeepSeek V4 Pro, and Gemini 3 Flash. We additionally included GPT-4o as a lower-cost comparison model that has been widely-used in annotation tasks. All model outputs were evaluated using the same structural graph evaluation protocol.

\subsection{Results}
We evaluate the model performance by applying the graph structural evaluation method (Section \ref{sec:graph-eval}) between the gold graph and the model-predicted graphs. The performance of our production model is reported in Table~\ref{tab:edge_evaluation_combined}.

\begin{table*}[ht]
\centering
\small
\setlength{\tabcolsep}{5pt}
\renewcommand{\arraystretch}{1.15}
\resizebox{\textwidth}{!}{%
\begin{tabular}{lccccccccc}
\toprule
\textbf{Type}
& \multicolumn{6}{c}{\textbf{Micro / Pooled Edges}}
& \multicolumn{3}{c}{\textbf{Macro / Per Abstract}} \\
\cmidrule(lr){2-7}
\cmidrule(lr){8-10}
& \textbf{Human}
& \textbf{LLM}
& \textbf{Matched}
& \textbf{Precision}
& \textbf{Recall}
& \textbf{F1}
& \textbf{Mean Precision}
& \textbf{Mean Recall}
& \textbf{Mean F1} \\
\midrule
Mechanistic
& 759 & 757 & 583 & 0.770 & \textbf{0.768} & \textbf{0.769}
& \textbf{0.829} & \textbf{0.826} & \textbf{0.798}  \\

Associational
& 282 & 297 & 206 & 0.694 & 0.730 & 0.712
& 0.725 & 0.742 & 0.711 \\

Moderational
& 253 & 187 & 151 & \textbf{0.807} & 0.597 & 0.686
& 0.728 & 0.619 & 0.639 \\

Hierarchical
& 418 & 417 & 272 & 0.652 & 0.651 & 0.651
& 0.690 & 0.668 & 0.662 \\

\midrule
Higher-Level Graph
& 768 & 800 & 570 & 0.713 & 0.742 & 0.727
& 0.743 & 0.789 & 0.732  \\

Type-Agnostic Graph
& 1712 & 1658 & 1261 & 0.761 & 0.737 & 0.748
& 0.813 & 0.815 & 0.779  \\

Full Graph
& 1712 & 1658 & 1212 & 0.731 & 0.708 & 0.719
& 0.767 & 0.771 & 0.736 \\

\bottomrule
\end{tabular}%
}
\caption{Structural evaluation of LLM-generated graphs against human-annotated gold graphs. Micro metrics are computed by pooling all edges across abstracts; we aggregate the total number of human-coded edges, LLM-predicted edges, and matched edges, and then compute precision, recall, and F1 from these corpus-level counts. Macro metrics are computed at the abstract level: precision, recall, and F1 are first calculated separately for each abstract and then averaged across abstracts.}
\label{tab:edge_evaluation_combined}
\end{table*}

We produced both micro and macro F1 scores for the best models (GPT-5.4 + GPT-5.2). Micro F1 is produced by pooling the edges and then computed as a whole, whereas macro F1 is computed by averaging the F1 over each abstract. Our best model (GPT-5.4 for Step 1, 5; GPT-5.2 for the rest) achieved a micro F1 of $0.72$ and a macro F1 of $0.74$. This is only slightly lower than our mean F1 score reported in inter-annotator agreement by human annotators. Our model performs well on extracting mechanistic and associational relationships. However, it performs less satisfactorily on extracting moderational and hierarchical relationships. Notably, moderational and hierarchical edges are intrinsically harder: the former requires higher-order reasoning beyond binary relation extraction, while the latter requires implicit taxonomic abstraction over variable mentions \citep{Luan_2018_Multi_Task_Identification, jia-etal-2019-document}.

We further assessed the robustness of model performance across journals and publish periods, see Appendix~\ref{tab:journal_edge_scores}. This suggests that the model generalizes well across abstracts from different historical periods, despite potential changes in writing style, reporting conventions, and terminology over time. Extraction performance is highly stable across period, with all F1 score above $0.71$. The performance varied more across journals, with journal-level F1 ranging from $0.67$ to $0.81$.

\subsection{Model Comparison}
Table~\ref{tab:model_performance} reports model performance under direct prompting and the proposed pipeline. The direct-prompting baseline obtains balanced but relatively low scores, with an F1 of $0.530$. In contrast, all pipeline-based configurations outperform the baseline. The combined GPT-5.2 + GPT-5.4 configuration achieves the best overall performance, with $0.767$ precision, $0.771$ recall, and $0.736$ F1. Among single-model configurations, GPT-5.4 performs best, reaching an F1 of $0.694$, followed by GPT-5.2 with an F1 of $0.679$.

Different models exhibit distinct error profiles. Gemini 3 Flash achieves the highest recall ($0.782$) but substantially lower precision, suggesting a tendency to over-generate relations. By contrast, DeepSeek V4 Pro and GPT-4o are more precision-oriented but recover fewer correct relations. We further compare direct prompting, chain-of-thought prompting, and the final staged pipeline using GPT-5.4. This comparison shows a clear improvement from direct prompting to chain-of-thought prompting and then to the staged pipeline, indicating that explicit decomposition improves graph extraction quality beyond model choice alone. Overall, the results show that the proposed pipeline improves the extraction relative to one-step prompting, and that combining GPT-5.2 and GPT-5.4 provides the best precision--recall balance.

\begin{table}[ht]
\centering
\small
\begin{tabular}{lccc}
\toprule
\textbf{Model} & \textbf{Precision} & \textbf{Recall} & \textbf{F1} \\
\midrule
GPT-5.4 (Direct Prompt)       & 0.566   & 0.606   & 0.528   \\
\midrule
GPT-5.2 + GPT-5.4 & \textbf{0.767} & 0.771 & \textbf{0.736 }\\
GPT-5.4            & 0.754 & 0.714 & 0.694 \\
GPT-5.4 (Single Step CoT) & 0.691 & 0.698 & 0.658\\
GPT-5.2            & 0.700 & 0.726 & 0.679 \\
GPT-4o             & 0.738 & 0.508 & 0.551 \\
Claude Opus 4.7       & 0.700 & 0.651 & 0.635 \\
Claude Sonnet 4.6     & 0.660 & 0.736 & 0.652 \\
Gemini 3 Flash     & 0.536 & \textbf{0.782} & 0.598 \\
Deepseek V4 Pro     & 0.757 & 0.619 & 0.641 \\
\bottomrule
\end{tabular}
\textit{Note. The model variants are gpt-4o, gpt-5.2, gpt-5.4, claude-sonnet-4.6, claude-opus-4.7, gemini-3-flash, deepseek-v4-pro accordingly. All parameters (i.e., reasoning level, verbosity) are set to 'low' where applicable.} 
\caption{Performance (macro-averaged) comparison of different models with respect to the gold graph.}
\label{tab:model_performance}
\end{table}

\subsection{Error Analysis}
\begin{figure}[ht]
    \centering
    \includegraphics[width=1\linewidth]{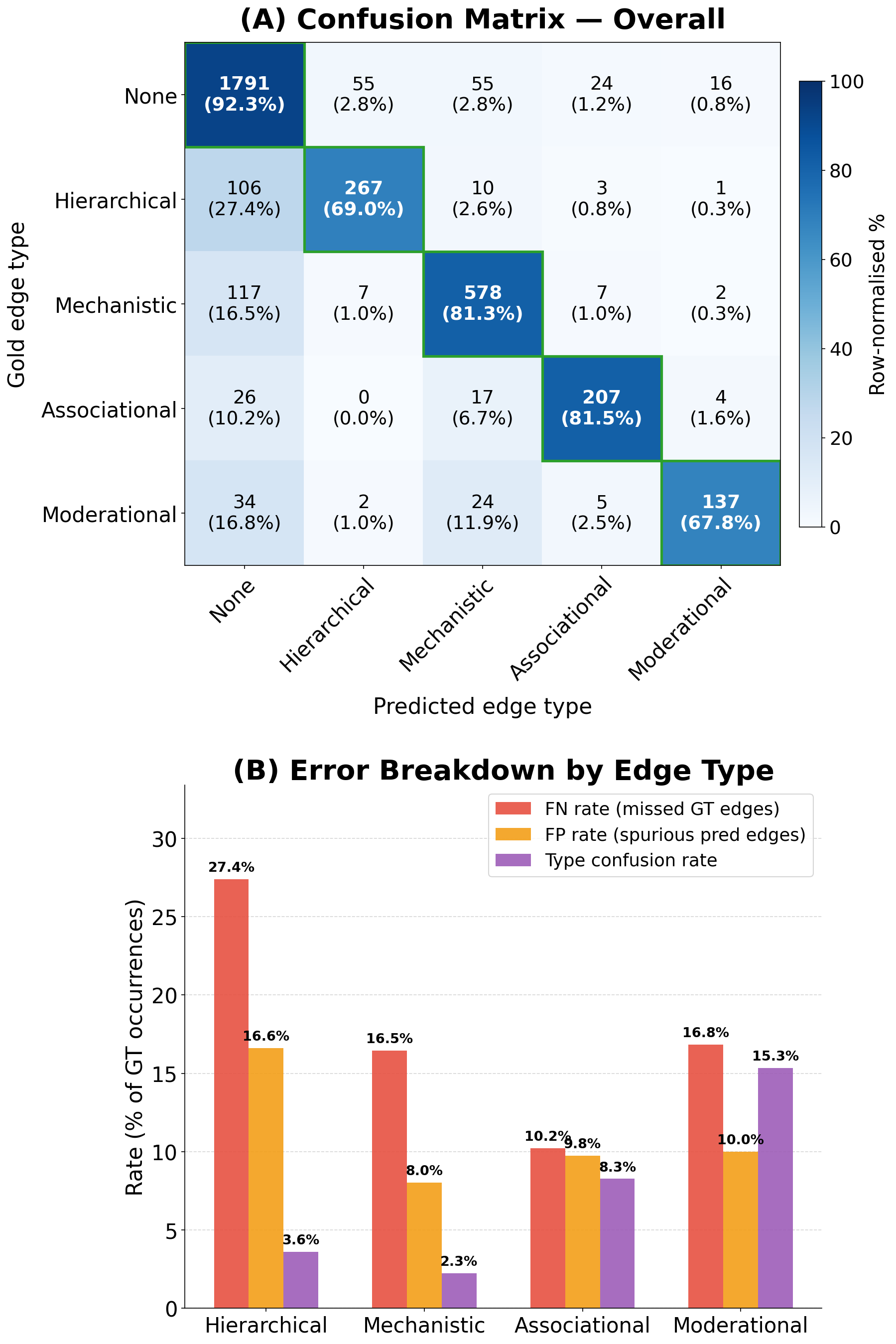}
    \caption{Confusion matrix of LLM-predicted edge types against gold graph edge types and error breakdown by edge types.}
    \label{fig:confusion}
\end{figure}

In addition to Table \ref{tab:edge_evaluation_combined}, Figure \ref{fig:confusion} presents a detailed confusion matrix and breakdown between the gold and predicted edge types. Hierarchy has the highest FN rate (27.4\%) — the model misses more than one in four hierarchy edges. This edge type requires inferring that one construct is a component of another, a structural inference rarely made explicit in abstracts.
Moderation has the highest type confusion rate (15.3\%) — nearly one-sixth of moderation edges are recovered under the wrong type, most often as directional. This reflects the inherent ambiguity of moderation language and the model's tendency to simplify three-way interactions.
Directional has the lowest type confusion (2.3\%) and a moderate FN rate, reflecting its dominant frequency and clearer linguistic markers (e.g., "predicted", "mediated", "caused").
Correlational edges are nearly balanced in FN/FP (10.2\% vs. 9.8\%), suggesting the model neither systematically over- nor under-produces correlational relationships, but does occasionally impose directionality (8.3\% type confusion).
FP rates are uniformly lower than FN rates across all types, indicating the model is more likely to miss a relationship than to hallucinate one—a desirable conservative bias for downstream knowledge-graph construction.

\section{Conclusion}

This study introduced EmpriGraph-Psy, a dataset and LLM pipeline for extracting empirical relation graphs from psychology abstracts. We explored how variable-centered scientific findings can be represented as graphs. We further demonstrated that decomposing graph construction into variable extraction, normalization, evidence selection, relation extraction, and validation substantially improves LLM performance over direct prompting. The best pipeline achieved strong graph-level extraction performance ($F1=0.74$), approaching human agreement and showing stable results across publication periods. EmpiriGraph-Psy provides a benchmark and extraction framework for constructing empirical knowledge graphs in psychology, and offers a reference framework for information extraction in broader variable-oriented empirical fields.

\section*{Limitations}
The current dataset is limited to psychology abstracts. Although the dataset covers multiple psychology subfields and a broad historical period, it remains unclear whether the proposed workflow generalizes to other disciplines such as health science or biology, where abstracts may follow different writing conventions and reporting styles. Future work could extend the workflow of Empirical Research Knowledge Graph Extraction to additional domains to assess the cross-disciplinary robustness of the proposed pipeline.

Our work focuses on extracting empirical and conceptual relationships and evaluating whether LLM-based pipelines can recover such graph structures from abstracts. The current annotation scheme does not include other important scientific components such as samples, methods, statistical procedures, or tasks. Future work may therefore integrate our LLM workflow with prior NLP-based scientific information extraction methods to construct more complete scientific knowledge graphs. 

\section*{Acknowledgments}
This work is funded by an UK ESRC Grant to the University of Warwick Centre for Competitive Advantage in the Global Economy (CAGE). XT is supported by the EPSRC [grant number EP/Y009800/1], through funding from Responsible AI UK (KP0016) as a Keystone project.

\bibliography{custom}

\appendix

\clearpage

\section{Appendix}
\label{sec:appendix}
\subsection{Dataset and Human Annotations}
\label{sec:appendix1}
\paragraph{Dataset} The final annotated dataset included abstracts from the following six journals: \textit{Journal of Applied Psychology}, \textit{Journal of Consulting and Clinical Psychology}, \textit{Journal of Counseling Psychology}, \textit{Journal of Educational Psychology}, \textit{Journal of Experimental Psychology: General}, and \textit{Behaviour Research and Therapy}. Our inclusion criteria were: (1) peer-reviewed journal articles and (2) empirical studies. Therefore, literature reviews, letters, meta-analyses, and other non-empirical studies were excluded from the dataset. For each journal, we sampled five abstracts from every 10-year period between the 1960s and 2025. The final dataset therefore consisted of 210 abstracts in total, including 30 abstracts per decade and 35 abstracts from each journal.

\begin{figure*}[ht]
    \centering
    \includegraphics[width=\linewidth]{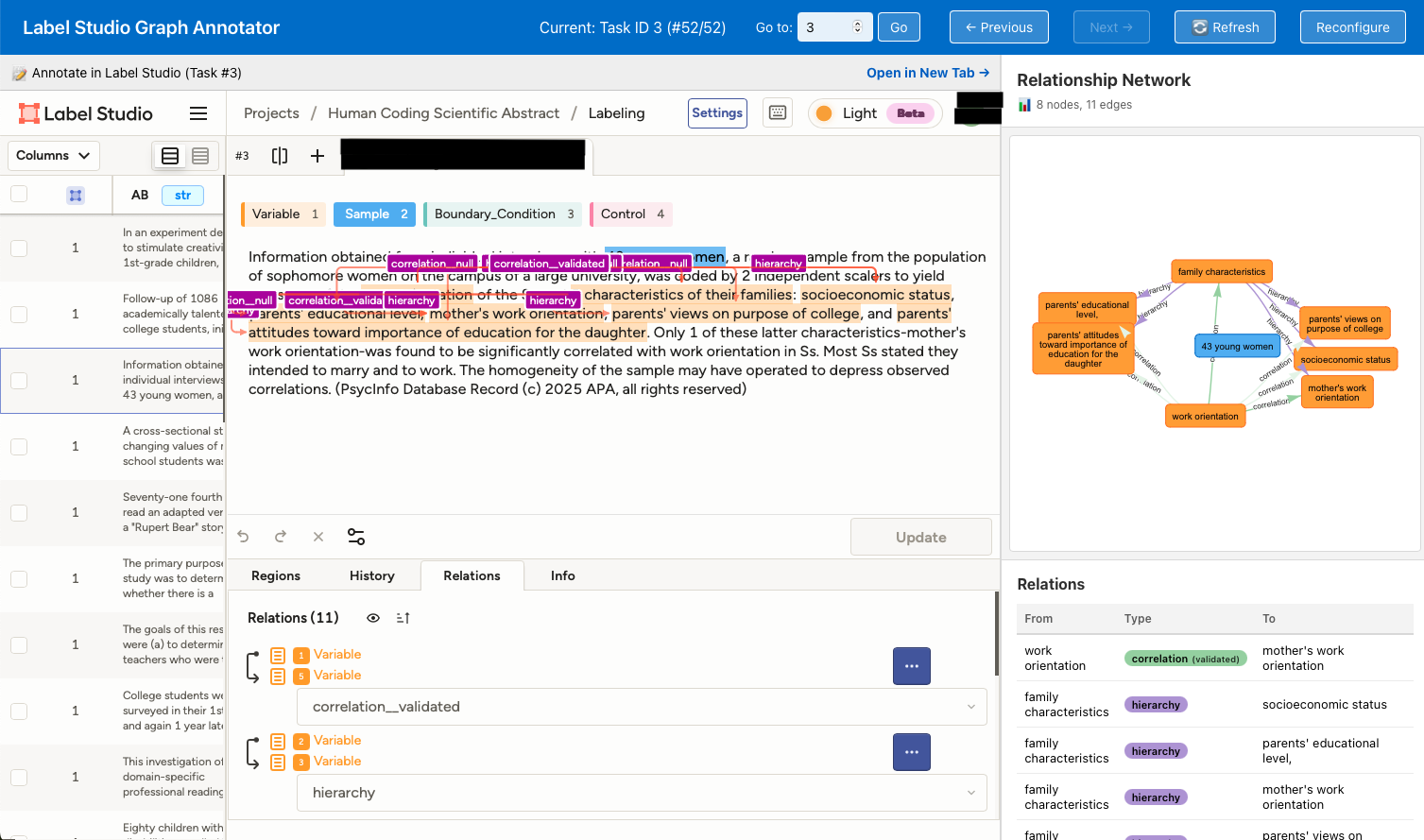}
    \caption{The user interface of annotation task}
    \label{fig:annotation_platform}
\end{figure*}

\paragraph{Human Annotation} Three annotators participated in the project. One coder (A) was an undergraduate student in Psychology, and the other two coders (B, C) were PhD students in Psychology. Annotators participated under different project arrangements, including compensated research assistance and unpaid project participation. The annotating interface can be found in Figure ~\ref{fig:annotation_platform}. The annotation interface extended Label Studio with custom modules for relational graph annotation. Annotators could highlight variable spans, modify normalized variable names in the region panel, assign relation types in the relations panel, and visually inspect the resulting relationship graph rendered in the sidebar after submission.

\paragraph{Annotation Guideline} 
Annotators followed a detailed annotation guideline. The guideline instructed human annotators to proceed through the following steps: (1) locate sentences in the abstract that contain information about key concepts, i.e., variables, and empirical relationships tested or reported in the article; (2) identify variables that were empirically examined in the study; (3) code hierarchical relationships between variables when one variable was presented as a subtype, component, dimension, indicator, or more specific measurement of a broader construct; (4) annotate the four relation types; (5) assign a validation-state label to each empirical relation, using \textit{validated}, \textit{null}, or \textit{hypothesized}; and (6) canonicalize variable names by merging synonyms and coreferential mentions. The full annotation guideline is provided in the anonymous repository linked in the abstract.

\paragraph{Inter-annotator Agreement} 

As reported in Table~\ref{tab:coder_agreement}, pairwise agreement was highest between Coder A and Coder C, both of whom were Psychology PhD students, with an F1 score of .82 and Cohen's $\kappa$ of .60. Comparisons involving Coder B, the Psychology undergraduate student, showed slightly lower but broadly comparable agreement.

\begin{table}[ht]
\centering
\small
\setlength{\tabcolsep}{8pt}
\renewcommand{\arraystretch}{1.12}
\begin{tabular}{lcc}
\toprule
\textbf{Coder comparison} & \textbf{F1} & \textbf{Kappa} \\
\midrule
Coder A vs. Coder B & 0.717 & Cohen's $\kappa = 0.566$ \\
Coder A vs. Coder C & 0.777 & Cohen's $\kappa = 0.590$ \\
Coder B vs. Coder C & 0.830 & Cohen's $\kappa = 0.559$ \\
\midrule
All three coders & \textemdash & Fleiss' $\kappa = 0.632$ \\
\bottomrule
\end{tabular}
\caption{Inter-annotator agreement among the three human coders on the 50-article overlap set. Pairwise F1 treats the second coder in each comparison as the reference annotation.}
\label{tab:coder_agreement}
\end{table}

To remain compatible with the access conditions of the original source materials, the released dataset does not redistribute copyrighted abstract text. Instead, we release metadata identifiers and the derived annotation layer, including normalized variables, empirical relation edges, hierarchy edges, validation states, and dataset splits. The annotation layer is released under a research-compatible open license, such as CC BY 4.0, while accompanying code is released separately under an open-source software license. The intended use of the dataset is research on scientific information extraction, empirical relation extraction, knowledge graph construction, and model evaluation. The dataset is not intended for clinical, legal, policy, or individual-level decision-making, nor for making evaluative claims about specific authors, participants, journals, or institutions.

\subsection{Model Performance Across Journals and Periods}

We assessed the robustness of model performance across journals and publish periods. As shown in Tables~\ref{tab:year_edge_scores}, extraction performance is highly stable across period, with all F1 score above $0.71$. This suggests that the model generalizes well across abstracts from different historical periods, despite potential changes in writing style, reporting conventions, and terminology over time. The performance varied more across journals, see Tables~\ref{tab:journal_edge_scores}, with journal-level F1 ranging from $0.67$ to $0.81$.

The strongest performance was observed for \textit{Journal of Consulting and Clinical Psychology} (JCCP; $F1=0.807$), which also showed the highest precision ($0.863$) and high recall ($0.806$). 
In contrast, performance was lowest for \textit{Behaviour Research and Therapy} (BRT; $F1=0.669$) and \textit{Journal of Experimental Psychology: General} (JEP:G; $F1=0.694$). We additionally examined the average number of edges and the distribution of edge types across journals. Lower-performing journals did not systematically contain more edges, nor did they contain a higher proportion of difficult edge types, such as moderational or hierarchical edges. This suggests that differences in journal-level performance are unlikely to arise from differences in graph complexity or relation-type composition. Instead, the variation is more likely attributable to journal-specific reporting styles, such as differences in how explicitly relations are stated or how consistently variables are described.
\begin{figure}[ht]
    \centering
    \includegraphics[width=1\linewidth]{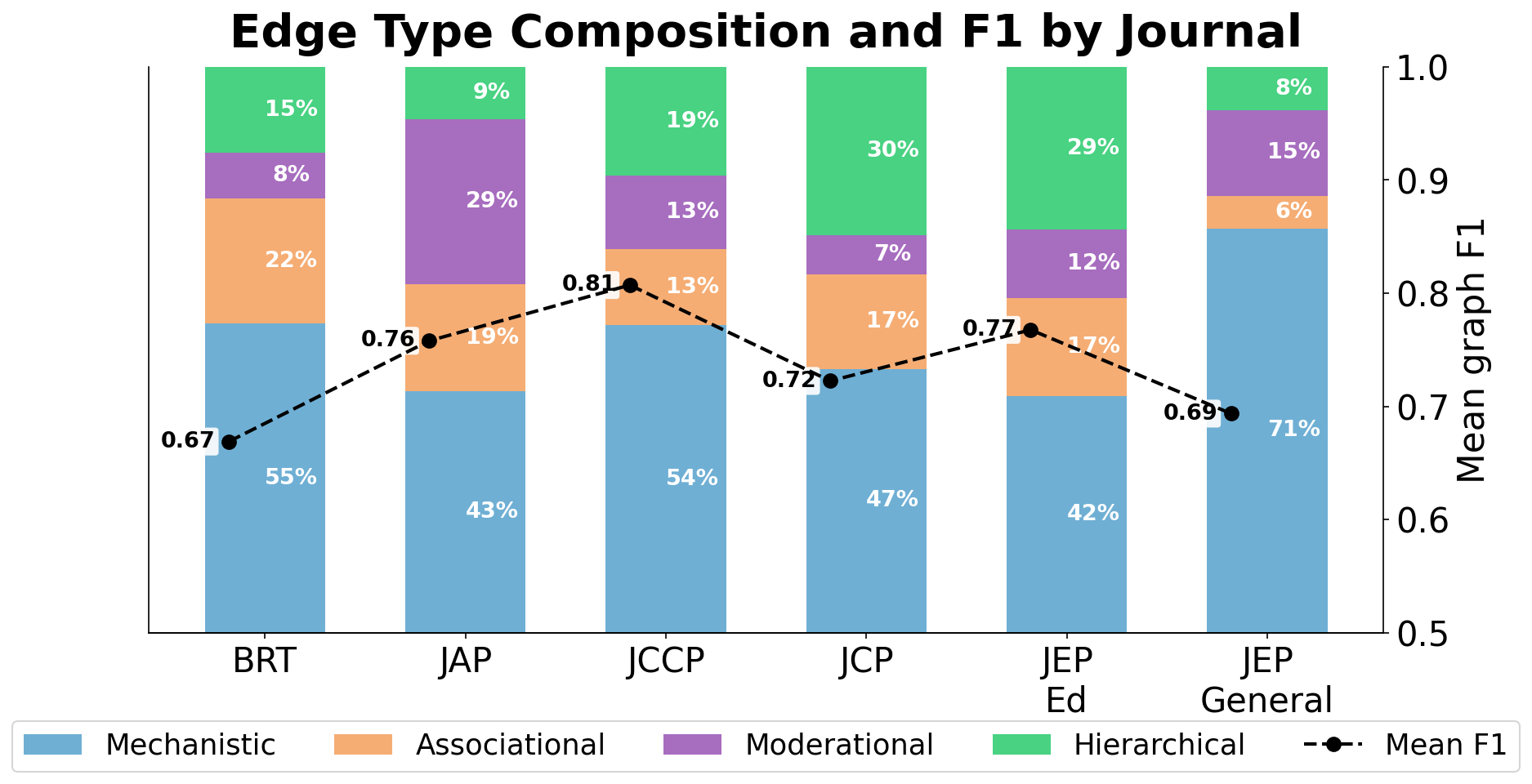}
    \caption{Edge Type Composition by Journal}
    \label{fig:placeholder}
\end{figure}

Overall, the results indicate that the proposed extraction pipeline is temporally robust, with consistently strong performance across publication periods, while journal-level variation remains a more important source of performance heterogeneity.

\begin{table}[ht]
\centering
\small
\begin{tabular}{lcccc}
\toprule
\textbf{Year Range} 
& \textbf{Abstracts}
& \textbf{Precision} 
& \textbf{Recall} 
& \textbf{F1} \\
\midrule
1960--1970 & 30 & 0.818 & 0.782 & 0.789 \\
1970--1980 & 30 & 0.717 & 0.804 & 0.718 \\
1980--1990 & 30 & 0.763 & 0.751 & 0.721 \\
1990--2000 & 30 & 0.769 & 0.719 & 0.710 \\
2000--2010 & 30 & 0.744 & 0.809 & 0.745 \\
2010--2020 & 30 & 0.777 & 0.782 & 0.743 \\
2020--2025 & 30 & 0.779 & 0.749 & 0.729 \\
\bottomrule
\end{tabular}
\caption{Average edge-level precision, recall, and F1 scores across year ranges.}
\label{tab:year_edge_scores}
\end{table}

\begin{table}[ht]
\centering
\small
\begin{tabular}{lcccc}
\toprule
\textbf{Journal} 
& \textbf{Abstracts} 
& \textbf{Precision} 
& \textbf{Recall} 
& \textbf{F1} \\
\midrule
BRT 
& 35 & 0.716 & 0.695 & 0.669 \\
JAP
& 35 & 0.798 & 0.749 & 0.758 \\
JCCP
& 35 & 0.863 & 0.806 & 0.807 \\
JCP
& 35 & 0.753 & 0.765 & 0.723 \\
JEP
& 35 & 0.760 & 0.818 & 0.768 \\
JEP:G
& 35 & 0.711 & 0.791 & 0.694 \\
\bottomrule
\end{tabular}
\textit{Note. BRT = Behaviour Research and Therapy; JAP = Journal of Applied Psychology; JCCP = Journal of Consulting and Clinical Psychology; JCP = Journal of Counseling Psychology; JEP = Journal of Educational Psychology; JEP:G = Journal of Experimental Psychology: General.} 
\caption{Average edge-level precision, recall, and F1 scores across journals.}
\label{tab:journal_edge_scores}

\end{table}

\subsection{Graph Evaluation Details}
\label{app:graph-eval}

We evaluate each predicted graph against a gold graph in two stages: node alignment followed by edge scoring under multiple evaluation views.

\paragraph{Preprocessing.}
After the final complete graphs are formed in Step 5, we apply a set of preprocessing rules before evaluation. These rules programmatically propagate relationships from lower-level to higher-level nodes and remove duplicate relationships. This improves consistency across annotators and models, and reduces errors caused by the complexity of manual annotation.

\paragraph{Graph representation.}
Let the gold graph be \(G=(V_G,E_G)\) and the predicted graph be \(P=(V_P,E_P)\). Edges are directed and typed:
\begin{equation}
e=(u,v,\tau), \qquad \tau \in \mathcal{T},
\end{equation}
where
\begin{equation}
\begin{aligned}
\mathcal{T}=\{&
\texttt{hierarchy}, \texttt{directional},\\
&\texttt{correlational}, \texttt{moderation}\}.
\end{aligned}
\end{equation}

\paragraph{Node alignment objective.}
We search for a partial mapping
\begin{equation}
\phi: V_G \to V_P \cup \{\emptyset\},
\end{equation}
where \(\emptyset\) indicates an unmatched gold node. The mapping is injective over non-empty assignments:
\begin{equation}
\phi(u)=\phi(v)\neq \emptyset \Rightarrow u=v.
\end{equation}
Thus, no two gold nodes can be matched to the same predicted node, while multiple gold nodes may remain unmatched.

A gold edge \((u,v,\tau)\in E_G\) is matched if and only if
\begin{equation}
\phi(u)\neq \emptyset,\qquad
\phi(v)\neq \emptyset,\qquad
(\phi(u),\phi(v),\tau)\in E_P.
\end{equation}
Define
\begin{equation}
M(\phi)=
\left|
\left\{
(u,v,\tau)\in E_G:
(\phi(u),\phi(v),\tau)\in E_P
\right\}
\right|.
\end{equation}
We optimize
\begin{equation}
\phi^\star=\arg\max_{\phi} M(\phi).
\end{equation}

This alignment is structure-first. Since the same variable may be realized in an abstract by multiple surface forms, annotators may not produce mutually consistent node labels unless the variable inventory is specified in advance. Under this mapping, nodes with different surface forms may nevertheless be aligned when their typed relational roles are compatible under \(\phi\). 

\paragraph{Search algorithm.}
Exact optimization of the alignment objective is combinatorial, so we use a branch-and-bound solver with a greedy warm start. First, we compute a greedy initial mapping \(\phi_0\), yielding an incumbent score \(B=M(\phi_0)\). We then run depth-first search over partial mappings \(\phi_d\).

At each search state, we compute the number of already matched edges \(C(\phi_d)\) and an admissible upper bound \(U(\phi_d)\) on the number of additional gold edges that could still be matched by any completion of \(\phi_d\). We use only safe pruning: a branch is pruned when
\begin{equation}
C(\phi_d)+U(\phi_d)\le B,
\end{equation}
because even the best possible completion of that branch cannot improve the incumbent. This pruning cannot remove any alignment that could achieve a higher score, provided that \(U(\phi_d)\) is an admissible upper bound.

In our implementation, \(U(\phi_d)\) is computed type-wise. For each relation type \(\tau\in\mathcal{T}\), let \(R_G^\tau(\phi_d)\) be the number of currently unmatched gold edges of type \(\tau\) whose endpoints are not ruled out by the partial mapping, and let \(R_P^\tau(\phi_d)\) be the number of available predicted edges of type \(\tau\) that could still be used by unmapped or consistently mapped endpoints. The number of additional matches of type \(\tau\) cannot exceed
\begin{equation}
\min\!\left(R_G^\tau(\phi_d), R_P^\tau(\phi_d)\right).
\end{equation}
Thus, we use
\begin{equation}
U(\phi_d)=
\sum_{\tau\in\mathcal{T}}
\min\!\left(R_G^\tau(\phi_d), R_P^\tau(\phi_d)\right).
\end{equation}
This bound is conservative: it may overestimate the number of additional matches because it ignores some joint compatibility constraints, but it never underestimates the maximum number of matches achievable by a completion. It is therefore safe for pruning.

Whenever a complete mapping with a higher score is found, the incumbent \(B\) and the best mapping are updated. Because search efficiency depends on node order, we run several ordering strategies, including degree-descending, degree-ascending, label order, and seeded random orders, and retain the best incumbent found:
\begin{equation}
M^\star = \max_{s\in S} M_s,
\end{equation}
where \(M_s\) is the best score found under strategy \(s\).

A timeout is enforced per strategy. If any strategy terminates exhaustively, its result is optimal under the specified objective, since the ordering affects only the search order and not the search space. Under timeout, the solver is anytime: it returns the best incumbent found so far, and the resulting alignment score is a lower bound on the optimal alignment score.

\begin{algorithm}[t]
\caption{Safe branch-and-bound alignment for typed directed graphs}
\label{alg:safe-bnb-alignment}
\begin{algorithmic}[1]
\Require Gold graph \(G=(V_G,E_G)\), predicted graph \(P=(V_P,E_P)\), edge types \(\mathcal{T}\), node order strategy \(s\)
\Ensure Best partial mapping \(\phi^\star\) and matched-edge count \(M^\star\)

\State \(\phi_0 \gets \Call{GreedyAlign}{G,P,s}\)
\State \(B \gets M(\phi_0)\)
\State \(\phi^\star \gets \phi_0\)
\State \(O \gets \Call{OrderNodes}{V_G,s}\)

\Procedure{Search}{$d,\phi_d$}
    \State \(C \gets \Call{MatchedEdges}{\phi_d,E_G,E_P}\)
    \State \(U \gets \Call{UpperBound}{\phi_d,E_G,E_P,\mathcal{T}}\)
    \If{\(C + U \leq B\)}
        \State \Return \Comment{safe pruning}
    \EndIf

    \If{\(d = |O|\)}
        \If{\(C > B\)}
            \State \(B \gets C\)
            \State \(\phi^\star \gets \phi_d\)
        \EndIf
        \State \Return
    \EndIf

    \State \(u \gets O[d]\)
    \ForAll{\(x \in V_P\) not already used by \(\phi_d\)}
        \State \(\phi' \gets \phi_d \cup \{u \mapsto x\}\)
        \State \Call{Search}{$d+1,\phi'$}
    \EndFor

    \State \(\phi' \gets \phi_d \cup \{u \mapsto \emptyset\}\)
    \State \Call{Search}{$d+1,\phi'$}
\EndProcedure

\State \Call{Search}{$0,\emptyset$}
\State \Return \(\phi^\star, B\)
\end{algorithmic}
\end{algorithm}

\paragraph{Edge scoring.}
Let \(m_G=|E_G|\), \(m_P=|E_P|\), and \(m_\star=M(\phi^\star)\). We compute:
\begin{equation}
\mathrm{P}=\frac{m_\star}{m_P},\qquad
\mathrm{R}=\frac{m_\star}{m_G},\qquad
\mathrm{F1}=\frac{2\mathrm{P}\mathrm{R}}{\mathrm{P}+\mathrm{R}}.
\end{equation}
These are the complete-graph structural scores under the active evaluation view.

For typed evaluation, we also compute per-type scores. For each \(\tau\in\mathcal{T}\), define:
\begin{equation}
E_G^\tau=\{(u,v):(u,v,\tau)\in E_G\},
\end{equation}
\begin{equation}
E_P^\tau=\{(x,y):(x,y,\tau)\in E_P\},
\end{equation}
and
\begin{equation}
M_\tau=
\left|
\left\{
(u,v)\in E_G^\tau:
(\phi^\star(u),\phi^\star(v))\in E_P^\tau
\right\}
\right|.
\end{equation}
Then
\begin{equation}
\label{eq:F1}
\mathrm{P}_\tau=\frac{M_\tau}{|E_P^\tau|},\qquad
\mathrm{R}_\tau=\frac{M_\tau}{|E_G^\tau|},\qquad
\mathrm{F1}_\tau=
\frac{2\mathrm{P}_\tau\mathrm{R}_\tau}
{\mathrm{P}_\tau+\mathrm{R}_\tau}.
\end{equation}

\paragraph{Evaluation views.}
We report three complementary alignment scores:
\(M_{\text{typed}}\), computed on the full graphs with directed, typed edges;
\(M_{\text{higher}}\), computed after projecting both graphs onto higher-level nodes by removing lower-level hierarchy children and hierarchy edges; and
\(M_{\text{agnostic}}\), computed on the full graphs after collapsing all edge types into a single undirected relation type.

\paragraph{Positive-edge scoring.}
For validated-only analysis, define
\begin{equation}
E_G^+ = \{e\in E_G: \mathrm{val}(e)=\texttt{validated}\},
\end{equation}
\begin{equation}
E_P^+ = \{e\in E_P: \mathrm{val}(e)=\texttt{validated}\}.
\end{equation}
We then apply the same alignment and scoring procedure to \((V_G,E_G^+)\) and \((V_P,E_P^+)\).

\paragraph{Complexity.}
Let \(n=|V_G|\), \(m=|V_P|\), \(e_G=|E_G|\), \(e_P=|E_P|\), and \(K=|S|\). Preprocessing operations, including canonicalization, deduplication, type filtering, and hierarchy transforms, are near-linear in edge count per graph in practice.

For alignment, the worst-case complexity is combinatorial:
\begin{equation}
O\!\left(
\sum_{k=0}^{\min(n,m)}
\binom{n}{k}
\frac{m!}{(m-k)!}
\right).
\end{equation}
Branch-and-bound often reduces the number of explored states substantially, but it does not change the worst-case complexity class. Across \(K\) ordering strategies, the complexity is \(O(K\cdot T_{\text{search}})\), with \(T_{\text{search}}\) capped by the timeout per strategy.

Space per graph pair is \(O(n+m+e_G+e_P)\) for graph structures, adjacency, and hashed edge sets, plus \(O(n+m)\) search-state overhead. At corpus scale, with \(A\) independent graph pairs and \(W\) workers, wall-clock runtime is approximately
\begin{equation}
T_{\text{wall}} \approx
\frac{\sum_{i=1}^{A} T_i}{W}
+
\text{parallel overhead}.
\end{equation}

\paragraph{Implementation note.}
The method is related to maximum common subgraph and maximum common edge-subgraph formulations, which search for structurally consistent correspondences between graphs under injective node mappings
\citep{mccreesh2017partitioning,ndiaye2011cp}. However, our evaluator optimizes a task-specific objective: maximum typed-directed edge overlap under an injective partial node mapping. It is not a classical induced-MCS objective, because non-edges are not required to be preserved and extra predicted edges are penalized through precision rather than through the alignment constraint. The branch-and-bound solver is exact when search terminates exhaustively; under timeout, it behaves as an anytime solver and returns the best incumbent found.

\subsection{Algorithm for Inter-coder Agreement}
\paragraph{Pairwise Cohen's kappa.}
For two raters $A,B$, collect labels $\{y_A(i),y_B(i)\}_{i=1}^{N}$ across all aligned node-pair items. Let $n_{pq}$ be confusion counts over classes $p,q\in\mathcal{C}$, and $N=\sum_{p,q}n_{pq}$.
Observed agreement:
\begin{equation}
P_o=\frac{1}{N}\sum_{c\in\mathcal{C}} n_{cc}.
\end{equation}
Chance agreement from marginals:
\begin{equation}
P_e=\sum_{c\in\mathcal{C}}
\left(\frac{n_{c\cdot}}{N}\right)\left(\frac{n_{\cdot c}}{N}\right).
\end{equation}
Cohen's kappa:
\begin{equation}
\kappa=\frac{P_o-P_e}{1-P_e}.
\end{equation}
We report pairwise $\kappa$ for all rater pairs (e.g., $A$-$B$, $A$-Gold, $B$-Gold).
\paragraph{Three-rater Fleiss' kappa.}
For three-rater agreement, we use the gold graph as pivot:
compute $\phi_{G\to A}$ and $\phi_{G\to B}$, keep gold nodes mapped to both raters, then form undirected node pairs on those shared gold nodes. Each item has three labels $(y_G,y_A,y_B)$.
Let $m=3$ raters, $K=|\mathcal{C}|=5$, $N$ items, and $n_{ij}$ the number of raters assigning item $i$ to class $j$. Per-item agreement:
\begin{equation}
P_i=\frac{1}{m(m-1)}\sum_{j=1}^{K} n_{ij}(n_{ij}-1).
\end{equation}
Mean observed agreement:
\begin{equation}
\bar P=\frac{1}{N}\sum_{i=1}^{N}P_i.
\end{equation}
Category prevalence:
\begin{equation}
p_j=\frac{1}{Nm}\sum_{i=1}^{N}n_{ij}.
\end{equation}
Chance agreement:
\begin{equation}
\bar P_e=\sum_{j=1}^{K}p_j^2.
\end{equation}
Fleiss' kappa:
\begin{equation}
\kappa_F=\frac{\bar P-\bar P_e}{1-\bar P_e}.
\end{equation}

\subsubsection{Intercoder Agreement via MCS-Aligned Node-Pair Labels}
\label{sec:intercoder-kappa}
We compute intercoder agreement on graph annotations using a two-stage protocol: structural node alignment followed by multiclass kappa on aligned node pairs.
\paragraph{Why alignment is required.}
Coders may use different surface names or granularity for conceptually similar variables. Direct string-based matching would therefore underestimate agreement. To make coder labels comparable, we first align nodes structurally using the same MCS-style logic used in our graph structural evaluation pipeline.
\paragraph{Graphs and labels.}
For each article, each human coder provides a preprocessed directed typed graph $G=(V,E)$, with edge types in
\begin{equation}
\begin{aligned}
\mathcal{T}=\{\texttt{hierarchy},\texttt{directional},\\
\texttt{correlational},\texttt{moderation}\}.
\end{aligned}
\end{equation}
Preprocessing is identical to the main graph evaluation workflow (correlation canonicalization, hierarchy handling, deduplication, and relation-priority collapse).
\paragraph{MCS-style node alignment.}
Given two rater graphs $G_1=(V_1,E_1)$ and $G_2=(V_2,E_2)$, we estimate an injective mapping
\begin{equation}
\phi: V_1 \to V_2
\end{equation}
that maximizes typed directed edge consistency. For a candidate $\phi$,
\begin{equation}
M(\phi)=\left|\left\{(u,v,\tau)\in E_1:(\phi(u),\phi(v),\tau)\in E_2\right\}\right|.
\end{equation}
We use a greedy multi-order approximation: run greedy mapping under several node orders (degree-descending, degree-ascending, label order, seeded random orders), then select the mapping with the largest $M(\phi)$:
\begin{equation}
\phi^\star = \arg\max_{\phi \in \Phi_{\text{greedy}}} M(\phi).
\end{equation}
This reuses the same MCS-inspired structural matching principle as the main evaluator.
\paragraph{Agreement unit construction.}
After alignment, agreement is computed over \emph{ordered node pairs} among mapped nodes:
\begin{equation}
\mathcal{U}=\{(u,v):u,v\in \mathrm{dom}(\phi^\star),\ u\neq v\}.
\end{equation}
Each ordered pair $(u,v)\in\mathcal{U}$ is one agreement item. If
$m=|\mathrm{dom}(\phi^\star)|$, the article contributes
\begin{equation}
m(m-1)
\end{equation}
items. Using ordered pairs preserves directional orientation: the item $(u,v)$ is
distinct from $(v,u)$.

\paragraph{5-class directed pair label.}
Each ordered pair receives one class from
\begin{equation}
\begin{aligned}
\mathcal{C}=\{\texttt{none},\texttt{hierarchy},
\texttt{directional},\\
\texttt{correlational},\texttt{moderation}\}.
\end{aligned}
\end{equation}
For rater 1, the label for $(u,v)$ is determined by the directed edge from $u$
to $v$ in $G_1$. For rater 2, the corresponding label is determined by the
directed edge from $\phi^\star(u)$ to $\phi^\star(v)$ in $G_2$. If no edge is
present in the relevant direction, the label is \texttt{none}. If multiple
relation types are present for the same ordered pair, we assign a single label
by the following priority:
\begin{equation}
\begin{aligned}
\texttt{directional} \succ \texttt{correlational}
\succ \\
\texttt{moderation} \succ \texttt{hierarchy} \succ \texttt{none}.
\end{aligned}
\end{equation}
Thus, each rater contributes one nominal label per aligned ordered node pair,
and disagreements in edge direction are reflected as label disagreements.

\paragraph{Interpretation.}
Because kappa is computed after MCS-aligned node correspondence, it measures
agreement on relation coding decisions under structural correspondence rather
than agreement on exact variable strings. Since agreement units are ordered node
pairs, the statistic is sensitive to both relation type and directional
orientation.
\paragraph{Complexity.}
Let $n_1=|V_1|$, $n_2=|V_2|$, $e_1=|E_1|$, $e_2=|E_2|$, and $|S|$ node-order strategies.
Per article, alignment cost is approximately
\begin{equation}
O\!\left(|S|\cdot n_1 n_2 \cdot d\right),
\end{equation}
where $d$ is local edge-consistency check cost with hashed edge lookups. If
$m=|\mathrm{dom}(\phi^\star)|$, directed pair-label construction is
\begin{equation}
O(m(m-1))=O(m^2).
\end{equation}
Kappa computation over $N$ total directed-pair items and fixed $K=5$ classes is
linear in $N$. Thus corpus-level time is additive over articles:
\begin{equation}
\sum_{a=1}^{A}\left(T_{\text{align}}^{(a)}+O(m_a^2)\right),
\end{equation}
with memory dominated by graph structures, mappings, and label vectors:
\begin{equation}
O(|V|+|E|+N).
\end{equation}

\subsection{Node Validation}
\label{app:node-validation}

Because the graph-structural evaluation aligns nodes by maximizing matched typed directed edges rather than by directly comparing node labels, we conduct an additional validation of the node correspondences induced by the common graph. This analysis asks whether structurally aligned node pairs also tend to be semantically similar at the label level. If structurally aligned nodes were frequently semantically unrelated, the reported edge-level F1 could be inflated by graph-theoretic coincidences rather than reflecting meaningful relation extraction.

\begin{table}[ht]
\centering
\small
\begin{tabular}{lcc}
\toprule
\textbf{Cosine similarity tier} & \textbf{Count} & \textbf{Share} \\
\midrule
$\ge 0.9$ exact        & 444 & 37.9\% \\
0.7--0.9 high          & 256 & 21.9\% \\
0.5--0.7 moderate      & 228 & 19.5\% \\
0.3--0.5 weak          & 140 & 12.0\% \\
$<0.3$ low             & 103 & 8.8\% \\
\bottomrule
\end{tabular}
\caption{Distribution of embedding-based cosine similarities between structurally aligned gold and predicted node labels.}
\label{tab:node_embedding}
\end{table}

\begin{figure}[ht]
    \centering
    \includegraphics[width=1\linewidth]{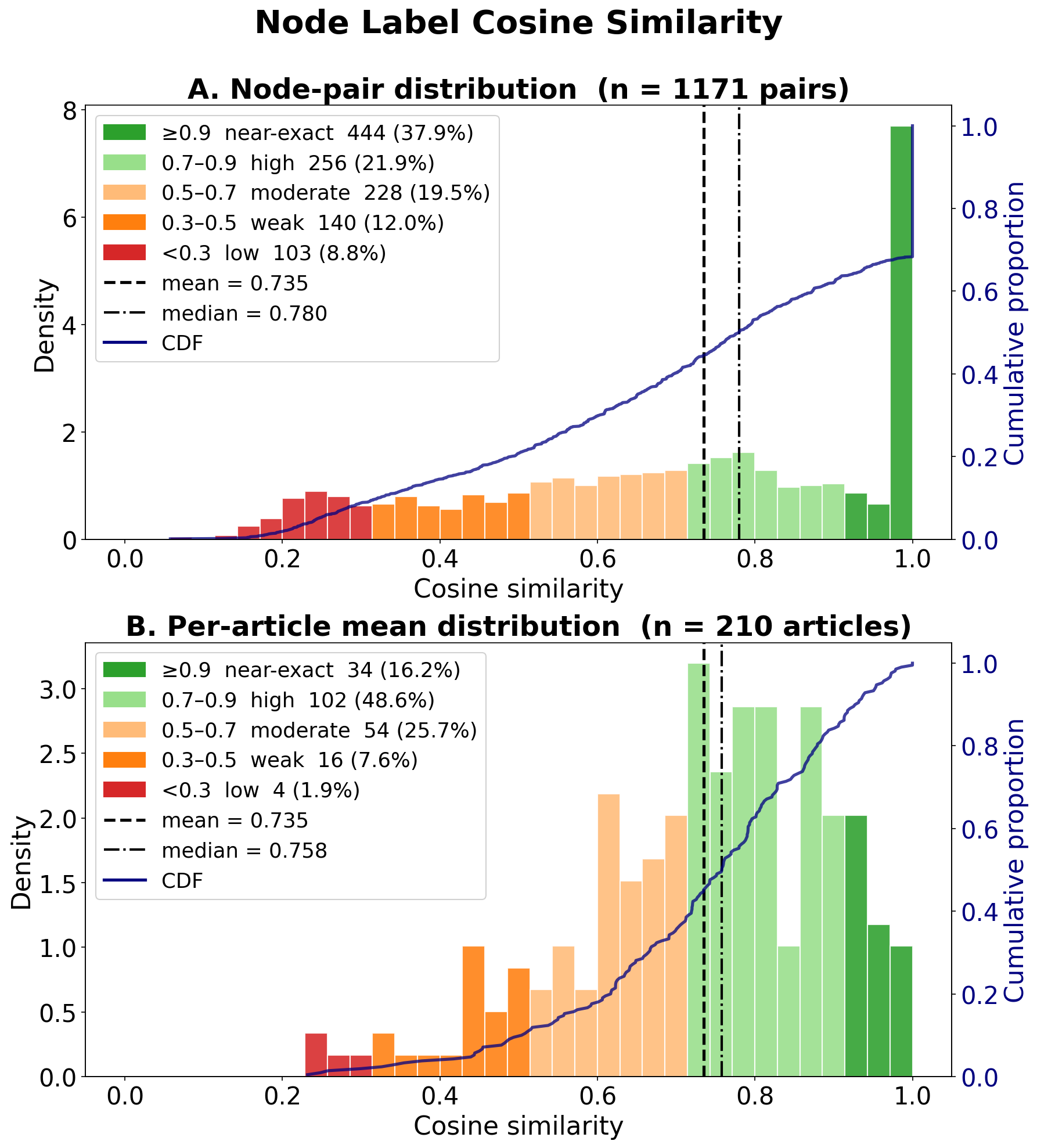}
    \caption{Distribution of cosine similarities between structurally aligned gold and predicted node labels.}
    \label{fig:node_cosine_distribution}
\end{figure}

Table~\ref{tab:node_embedding} and Figure~\ref{fig:node_cosine_distribution} show the cosine similarity distribution for gold--predicted node pairs appearing in the common graph. Across all aligned node pairs, the mean embedding-based cosine similarity is $0.735$. The majority of aligned pairs fall in the moderate-to-exact similarity range, suggesting that the MCS-derived alignment is usually pairing variables with related meanings rather than merely matching nodes with compatible graph positions.

We further stratify aligned node pairs by embedding-based cosine similarity and then randomly sample 100 pairs in proportion to the observed tier distribution. Human inspection of this sample assesses whether aligned nodes refer to the same empirical variable or construct despite surface-form differences. In this sample, $87$ pairs were judged to refer to the same variable or construct. Almost all cosine similarity $\ge 0.5$ pairs refer to the same construct. About one-third of the $0.3-0.5$ similarity tier node pairs are correctly paired; an example is 'general working model of attachment' and 'attachment variables,' yielding cosine similarity of $0.35$ but refer to the same variable in the abstract. All $<0.3$ sample pairs are incorrectly paired in reality. This result supports the validity of the typed-directed edge F1 metric: most matched edges are evaluated over semantically corresponding variable pairs, even though node-label similarity is not used as the primary alignment objective. 

\begin{table}[ht]
\centering
\small
\begin{tabular}{lc}
\toprule
\textbf{Edge Type} & \textbf{Mean Cosine Similarity}  \\
\midrule
Mechanistic        & 0.732 \\
Associational          & 0.752  \\
Moderational      & 0.764  \\
Hierarchical          & 0.727  \\
\bottomrule
\end{tabular}
\caption{Edge type cosine similarity derived by averaging node similarity from both sides of the edge.}
\label{tab:edge_embedding}
\end{table}

Table~\ref{tab:edge_embedding} reports edge-level cosine similarity by relation type, computed by averaging the cosine similarities of the two aligned node pairs forming each matched edge. The mean similarities are broadly comparable across relation types, ranging from $0.727$ for hierarchical edges to $0.764$ for moderational edges. This suggests that the structural alignment procedure produces semantically plausible node correspondences across all relation categories, rather than relying on a single relation type with unusually high label similarity. Hierarchical edges show the lowest mean similarity, which is expected because they often connect broader constructs with finer-grained variables or measurements.

Overall, the node-pair cosine analysis and manual validation provide a validity check for the structural evaluation procedure. Since the primary alignment objective does not use node-label similarity, the observed semantic similarity among common-graph node pairs suggests that the structural matching procedure usually recovers meaningful node correspondences. Therefore, the reported graph F1 can be interpreted as a relation-extraction score over largely semantically aligned variables, rather than as an artifact of arbitrary node matching.

\subsection{AI Assistance Statement}
The manuscript was written by the authors. ChatGPT (\url{https://chat.openai.com/}) was used only for language polishing. Claude Code was used for coding assistance such as debugging during implementation. Code and final manuscript were checked and verified by the authors.

\end{document}